\documentclass[review,sort&compress,3p]{elsarticle}
\usepackage[utf8]{inputenc}
\usepackage{makeidx}
\usepackage{graphicx,color}% Include figure files
\usepackage[dvipsnames]{xcolor}
\usepackage{gensymb}
\usepackage{geometry}
\usepackage{amsmath}
\usepackage[normalem]{ulem}
\usepackage{lineno,hyperref}
\usepackage{soul}
\begin{document}

\begin{frontmatter}

\title{Auxetic properties of a newly proposed $\gamma$-graphyne-like material}

\author[label1]{Ricardo Paupitz}
\author[label2]{Tales J. da Silva} 
\author[label2]{Marilia J. Caldas}
\author[label3]{Douglas S. Galvão}
\author[label3]{Alexandre F. Fonseca}
\address[label1]{Physics Department, São Paulo State University - UNESP, 13506-900, Rio Claro, SP, Brazil.}
\address[label2]{Institute of Physics, University of São Paulo, CEP 05508-090, São Paulo, SP, Brazil.}
\address[label3]{Applied Physics Department, Institute of Physics “Gleb Wataghin”, State University of Campinas, Campinas, SP, 13083-970, Brazil.}

\begin{abstract}
In this work, we propose a new auxetic (negative Poisson's ratio
values) structure, based on a $\gamma$-graphyne structure, here named
{\it A$\gamma$G structure}. Graphynes are 2D carbon allotropes with
phenylic rings connected by acetylenic groups. The A$\gamma$G
structural/mechanical and electronic properties, as well as its
thermal stability, were investigated using classical reactive and
quantum molecular dynamics simulations. We found that A$\gamma$G has a
large bandgap of 2.48 eV and is thermally stable at a large range of
temperatures. It presents a Young's modulus that is an order of
magnitude smaller than that of graphene or $\gamma$-graphyne.  The
classical and quantum results are consistent and validate that the
A$\gamma$G is auxetic, both when isolated (vacuum) and when deposited
on a copper substrate. We believe that this is the densest auxetic
structure belonging to the graphyne-like families.
\end{abstract}

\begin{keyword}
Carbon materials, Elastic properties, Simulation and modelling 

\end{keyword}

\end{frontmatter}

%%%MAIN TEXT%%%%
\section{Introduction}
Auxeticity is an interesting but unusual mechanical property only
present in few materials. Auxetic materials transversely expand
(contract) when longitudinally stretched
(compressed)~\cite{reviewEvans2000,review2Evans2000}. They are
characterized by negative values of the Poisson ratio, $\nu$, defined
as $\nu=-\epsilon_y/\epsilon_x$, where $\epsilon_y$ ($\epsilon_x$) is
the transverse (longitudinal) strain in response to the applied
longitudinal stress~\cite{neville2013}. Poisson ratio values are
usually positive while being negative for auxetic
materials~\cite{evans1991scicorrespondence,lakes1987,lakes1993,raydouglas1993,QingAuxetichoneycomb2019}. Zero,
or close to zero values are even more
rare~\cite{silvaCORK2005,grima2011,ray2015,fonseca2020}.

Auxeticity in nanomaterials has been object of intense
research~\cite{review2016,Mortazavi2017}. In particular, several
studies of molecular and/or polymer networks with negative and/or zero
Poisson's ratio are found in the
literature~\cite{raydouglas1993,grima2011,gibs1982,woj1987,evans1994,evans1995,griffin1998,grimaevans2000,grima2008,susuki2016,LiACSOmega2020}. As
the Poisson ratio is scale-independent, a structure possessing
negative or zero Poisson's ratio at molecular level is expected to
manifest this property at macroscopic scale.

Graphynes are two-dimensional one-atom-thick carbon allotropes. Their
structures can be formed by any possible combination of connected
acetylenic ($-$C$\equiv$C$-$) groups not only to each other but also
to phenylic rings and/or carbon-carbon bonds, thus generating
different families of
structures~\cite{Baughman876687,ivanovsREVIEW2013,liREVIEW2014}. Graphyne-based
structures present electronic~\cite{malko2008nonnullbandgapGYs},
thermal~\cite{wangthermoeletric2013GYs,cevikthermoelectric2014GYs,fonseca2017mestradosergio}
and
mechanical~\cite{fonseca2017mestradosergio,buheler2011GYsmech,nuno2021,QPengGraphynes2020}
properties that make them suitable for a series of different
applications.
 
Most of the graphyne-based structures have positive Poisson's ratio
values ~\cite{fonseca2017mestradosergio}, however, in the last
decades, some auxetic structures were reported in the
literature. Evans and
collaborators~\cite{evans1991scicorrespondence,grima2011,grima2008}
reported the auxeticity of reentrant poly-phenylacetylene networks,
called {\it reflexynes}. Grima and Evans~\cite{grimaevans2000} showed
that networks of acetylenic polytriangles with benzene rings at the
vertices, named as {\it polytriangles-$n$-ynes}, with $n$ being the
number of acetylenic linkages, are also auxetic for $n \geq 2$.
\begin{figure}[h]
\centering
\includegraphics[width=3.2 in, keepaspectratio]{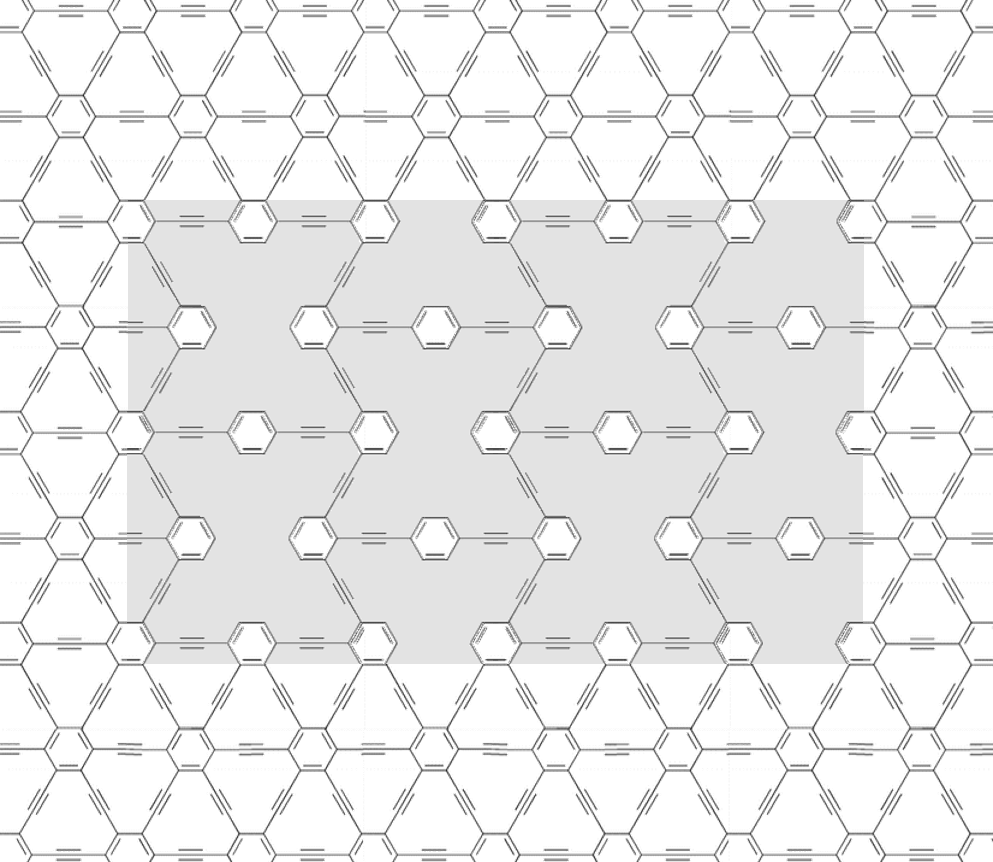}
\caption{The newly proposed auxetic $\gamma$-graphyne-like,
  (A$\gamma$G) is shown in the gray area. For reference, its parent
  regular $\gamma$-graphyne, in the non-gray area, is also shown. See
  text for discussions.}
\label{fig1}
\end{figure}

In this work, we report the design and investigation of the electronic
and mechanical properties of a newly proposed $\gamma$-graphyne-based
auxetic structure, that we name as {\it auxetic
  $\gamma$-graphyne-like} (A$\gamma$G), shown in the grey area of
Fig. \ref{fig1}. For reference, its parent $\gamma$-graphyne (non-grey
area) is also shown, in order to indicate how the auxetic structure
was obtained, which is by `deleting' specific acetylenic arms from a
regular $\gamma$-graphyne, thus creating a reentrant-like
structure. The resulting carbon atoms with dangling bonds are then
hydrogen passivated (for clarity, these hydrogen atoms are not shown
in Fig. \ref{fig1}). The elastic properties were obtained for the
infinite isolated A$\gamma$G layer, as well as for one finite
structure on a copper substrate. The thermal stability of this
structure was also tested to rule out the possibility of thermally
induced scrolling and/or folding.

\section{Computational Details}

Density Functional based Tight Binding (DFTB) and classical molecular
dynamics (MD) simulations were carried out to investigate the
structural stability, mechanical and thermal properties of the
A$\gamma$G. The geometry optimization of the structures was performed
by energy minimization through a conjugate-gradient algorithm
implemented in the LAMMPS code~\cite{plimpton1995}, with a force
tolerance set to $10^{-8}$ eV/\AA. Thermal stability was simulated
with a Langevin thermostat using a damping factor set to 1 ps.

The in-plane elastic properties of an infinite A$\gamma$G structure
was evaluated through~\cite{andrewPRB2019} the following equation:
\begin{equation}
\label{E2D}
    U(\varepsilon)=\frac{1}{2}C_{11}\varepsilon_{xx}^2+
    \frac{1}{2}C_{22}\varepsilon_{yy}^2+
    C_{12}\varepsilon_{xx}\varepsilon_{yy}\, ,
\end{equation}
where $U$ is the energy per area of the 2D structure, and
$\varepsilon_{xx}$ and $\varepsilon_{yy}$ are the strains along the
$x$ and $y$ directions, respectively. With the values of $C_{11}$,
$C_{12}$ and $C_{22}$, the Young's modulus, $Y$, the Poisson's ratio,
$\nu$, and the linear compressibility, $\beta$, along two in-plane
directions, can be calculated: $Y_{x} = \left(C_{11}C_{22} -
C_{12}^2\right)/C_{22}$, $Y_{y}=Y_{x}C_{22}/C_{11}$,
%$Y_{y} = \left(C_{11}C_{22} - C_{12}^2\right)/C_{11}$, 
$\nu_{xy} = C_{12}/C_{22}$, $\nu_{yx} = C_{12}/C_{11}$,  $\beta_x=\left(C_{22}-C_{12}\right)/K$ and $\beta_y=\left(C_{11}-C_{12}\right)/K$, with $K \equiv \left(C_{11}C_{22} - C_{12}^2\right)$.

$C_{ij}$, $i,j=1,2$, were obtained from two uniaxial tensile tests
(one along each direction) and one biaxial tensile test. The uniaxial
tensile test consists of pulling the structure along the $x$ ($y$)
direction while keeping its size along $y$ ($x$) direction fixed or
imposing $\varepsilon_{yy}=0$ ($\varepsilon_{xx}=0$). The biaxial test
consists of pulling the structure along the $x$ and $y$ directions at
the same time and by the same amount of strain
($\varepsilon_{xx}=\varepsilon_{yy}\equiv\varepsilon$). $C_{11}$
($C_{22}$) is obtained from an uniaxial tensile test, where
Eq. (\ref{E2D}) becomes $U(\varepsilon)=0.5C_{11}\varepsilon^2$
($U(\varepsilon)=0.5C_{22}\varepsilon^2$), that is further fitted for
$U$ versus uniaxial $\varepsilon$. $C_{12}$ is obtained from a biaxial
test, where Eq. (\ref{E2D}) becomes $U(\varepsilon)=M\varepsilon^2$,
where $M=C_{12}+0.5(C_{11}+C_{22})$, that is further fitted for $U$
versus biaxial $\varepsilon$.

Strain increments of 0.1\%, from 0 up to 1\%, were applied to the
A$\gamma$G structure under periodic boundary conditions, which is
consistent with the type of tensile test simulated. Energies per area,
$U$, were collected from the results of energy minimization as
described above, using three force-fields (AIREBO~\cite{brenner2002},
COMB3~\cite{tao2013} and ReaxFF~\cite{vanduin2001,mueller2010}) and
the DFTB method (described ahead) in order to verify if the mechanical
properties (and, in particular, the auxeticity of the structure) are
potential or method dependent. AIREBO, COMB3, and ReaxFF are three of
the most used MD potentials to the study of structural and mechanical
properties of carbon and other nanostructures. The ReaxFF parameters
used here are from Mueller {\it et al.}~\cite{mueller2010}. Fittings
of the data by a parabolic function on $\varepsilon$ were performed to
obtain $C_{ij}$, $i,j=1,2$ and, from them, the corresponding Young's
moduli, Poisson's ratio, and linear compressibility values.

The electronic structure calculations were carried within the DFTB
approximation, which can be described as a combination of Density
Functional Theory (DFT) and Tight Binding~\cite{dftb1,dftb2}
methods. This approximation allows quantum simulations of the
electronic and structural properties for relatively large systems
combining, as much as possible, the low computational cost of
tight-binding methods with the known precision of
DFT~\cite{Manzano123239}. In order to achieve this optimized
performance, DFTB is based on a second-order expansion of the
Kohn-Sham energy defined in DFT~\cite{koskinen2009density}. In the
present work we adopted the self consistent charge (SCC)-DFTB version,
as implemented in DFTB+ code~\cite{Elstner987260,dftb+} and the {\sl
  matsci}
parametrization~\cite{Elstner987260,Gemming2010,Lukose2010,Kubar2013}. For
the infinite structure cyclic boundary conditions were adopted in the
\textit{x,y} directions, with enough distance along \textit{z} to
avoid interaction between layers. Geometrical optimizations were
performed considering a conjugate gradient algorithm in which the
convergence for geometry was achieved with a maximum force difference
of 10$^{-6}$ and SCC iterations with a maximum tolerance of 10$^{-6}$.
Quantum Molecular Dynamics simulations for several temperatures were
performed with the Verlet algorithm and under the control of a
Nose-Hoover Thermostat~\cite{evans1985nose}, as implemented in DFTB+
software~\cite{dftb+}. Atomic structure visualizations and the
snapshots combined in the movies (see the Supplementary Materials)
were obtained with the VMD software~\cite{vmd}.

\section{Results and Discussion}

In Table \ref{table:elastic} we present the results for the elastic
constants, $C_{ij}$ (GPa.nm), Young's modulus, $Y$ (GPa.nm), Poisson's
ratio, $\nu$, and linear compressibility, $\beta$ (GPa.nm)$^{-1}$, of
the A$\gamma$G from the simulations with the MD potentials and
DFTB. All results are consistent with relation to the fact that
A$\gamma$G is auxetic, with Poisson's ratio values between -0.57 and
-0.86. For the Young's modulus of A$\gamma$G, DFTB values are among
the stiffest, with $Y_x=21.74$ (GPa.nm) being the highest value found,
while $Y_y=9.85$ (GPa.nm) obtained from COMB3 is the smallest. One
remarkable difference is that ReaxFF predicts symmetric behavior of
the elastic properties of A$\gamma$G, while AIREBO and COMB3 predict
asymmetric behavior. Also, results from DFTB predict asymmetric
behavior of the elastic properties, with absolute values closer to
those from the AIREBO potential.

\begin{table*}
    \centering
    \caption{Elastic constants $C_{ij}$ (GPa.nm), Young's modulus $Y$ (GPa.nm), Poisson's ratio $\nu$ \,and linear compressibility $\beta$ (GPa.nm)$^{-1}$ of the A$\gamma$G from different methods and MD potentials.}
    \begin{tabular}{cccccccccc}
    \hline
    \hline
     Method & $C_{11}$ & $C_{22}$ & $C_{12}$ & $Y_{x}$ & $Y_{y}$ & $\nu_{xy}$ & $\nu_{yx}$ & $\beta_x$ & $\beta_y$\\ 
    \hline
    MD AIREBO & 40.28 & 31.48 & -26.40 & 18.13 & 14.17 & -0.84 & -0.66 & 0.101 & 0.117 \\    
    MD COMB3 & 29.14 & 19.16 & -16.47 & 14.99 & 9.85 & -0.86 & -0.57 & 0.124 & 0.159 \\    
    MD ReaxFF & 31.16 & 32.60 & -23.00 & 14.93 & 15.62 & -0.71 & -0.74 & 0.114 & 0.111 \\
    DFTB  & 42.41 &  33.83 &  -26.44 & 21.74 & 17.34 & -0.78 & -0.62 & 0.082 & 0.094 \\
       \hline    
    \hline
    \end{tabular}
    \label{table:elastic}
\end{table*}

\begin{figure}[h]
\centering
\includegraphics[width=3.3 in, keepaspectratio]{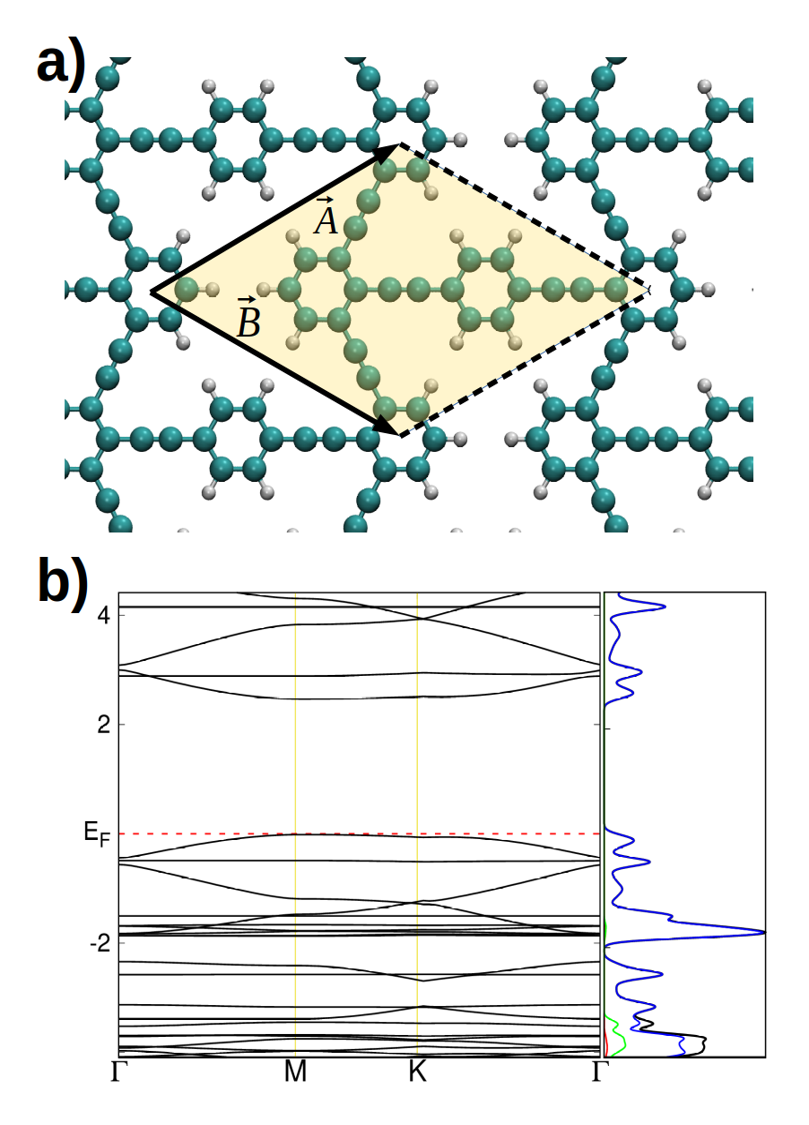}
\caption{a) The unit cell of the A$\gamma$G structure is indicated by
  the yellow area. The lattice vectors $\Vec{A}$ and $\Vec{B}$ have
  the same length ($\|\Vec{A}\|=\|\Vec{B}\|=12.12$ \AA) and forming a
  60$^{\degree}$ angle. Cyan and white spheres represent carbon and
  hydrogen atoms, respectively. b) DFTB band structure and DOS.
  $E_F$ is the Fermi level energy, highlighted by the red dashed
  line. Blue and green lines indicate the {\sl p} and {\sl s} orbital
  contributions from carbon atoms to the total density of states,
  while the red solid line shows the {\sl s} orbital contribution
  originated from the hydrogen atoms. The black line indicates the
  total density of states.}
\label{fig2}
\end{figure} 
The lattice vectors used to define the 2-dimensional Bravais lattice,
as well as the DFTB+ results for the density of states (DOS) and band
structure, are shown in Figure \ref{fig2}. The direct bandgap, around
2.48 eV, is located at the M point and suggests that the structure
would behave as an insulator or a wide bandgap semiconductor. The DOS
analysis confirms that, as expected, the top valence and lowest
conduction band states are formed by mainly contributions from the 2p
($\pi$) carbon orbitals. On the other hand, the small
dispersion/curvature observed at the bottom of the conduction band and
on the top of the valence band indicates charge carriers with
relatively large effective masses.

The thermal stability of A$\gamma$G structures was verified through MD simulations. The isolated structures (in vacuum) were simulated at room temperature using either classical or quantum MD methods. DFTB based MD simulations suggest that the structure remains stable at room temperature. For high temperatures, there were deformations but with no bond breaking up to 1000 K in vacuum conditions (videos S1 and S2 in the supplementary material). 

\begin{figure*}[h]
\centering
\includegraphics[width=6.3 in, keepaspectratio]{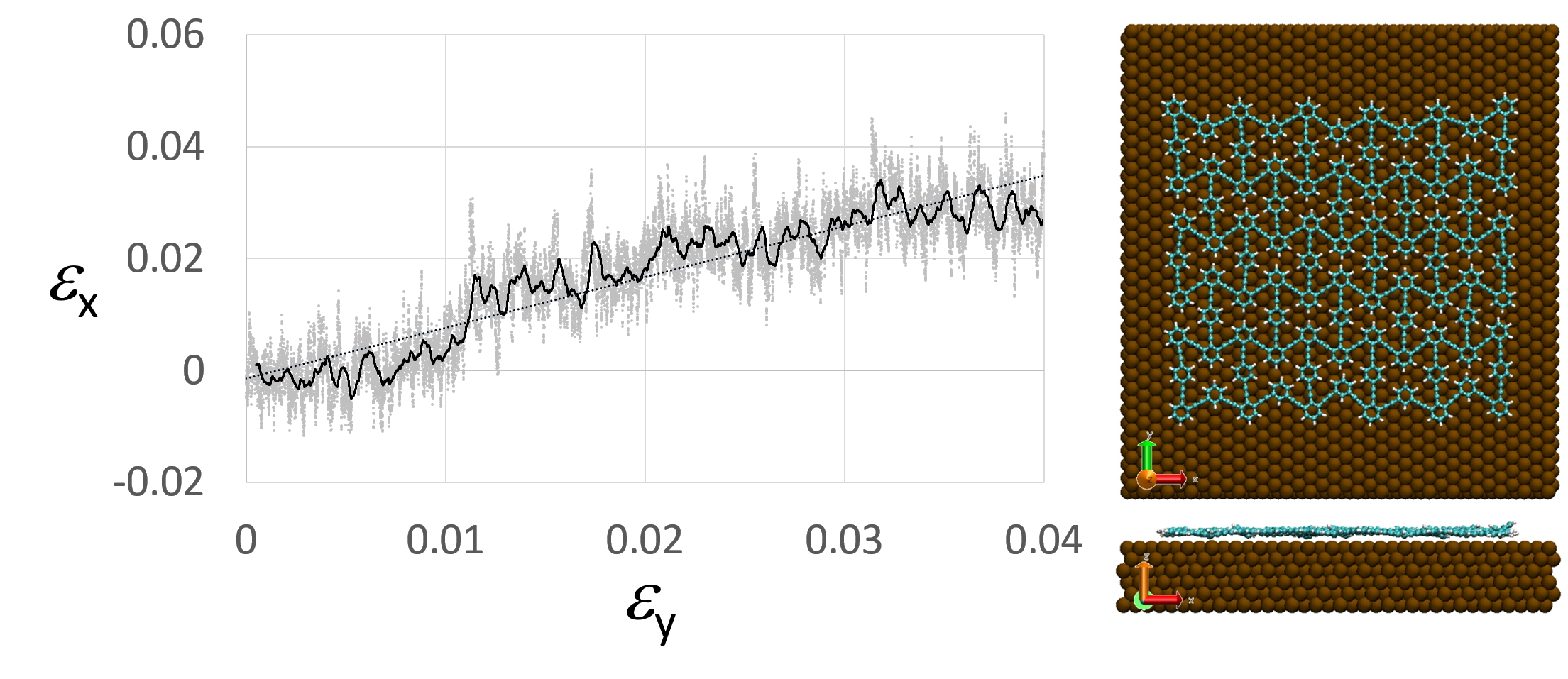}
\caption{Left panel: variation of the strain of the 63 $\times$ 60 \AA\mbox{ } A$\gamma$G structure on a copper substrate along the $x$ direction, as a function of the applied strain along the $y$ direction. Gray dots are raw data from the MD simulations. The black curve and black dotted line represent the average curve (made with 255 points), and the linear fitting of the data, respectively. Right panel: upper and lateral views of the structure. }
\label{fig3}
\end{figure*}

We have also carried classical MD simulations to test the thermal stability of a finite A$\gamma$G structure (of about 63 $\times$ 60 \AA\mbox{ }of size, see right panel of Fig. \ref{fig3}) deposited on a copper substrate. The structure is very stable at 300 K, but for high temperatures (about 1000 K), some atoms from the borders detach off the substrate after 200 ps (data not shown). 

Dynamical tensile strain tests were then simulated for the finite, hydrogen-passivated, A$\gamma$G structure, in order to verify its auxeticity at room temperature and deposited on the copper substrate (see right panel of Fig. \ref{fig3}). The COMB3 potential was used to simulate this configuration. Strain along the $x$ direction, $\varepsilon_x$, was collected as a function of the strain along the $y$ direction, $\varepsilon_y$, that was the direction of the applied tensile pulling stress. The carbon atoms at the bottom edge of the structure were fixed and the carbon atoms at the uppermost edge of the structure were moved along the $y$ direction at a strain rate of 0.01\% ps$^{-1}$. The results are shown in the left panel of Fig. \ref{fig3}. 

It is possible to see that $\varepsilon_x$ increases with $\varepsilon_y$, which is the signature of a negative Poisson's ratio and consequently of the auxetic behavior. The rate of increase of $\varepsilon_x$ is slow for low $\varepsilon_y$, but it suddenly increases. The average Poisson's ratio estimated for $\varepsilon_x = 4 \%$ is -0.91. The same test was simulated, now pulling the structure along the $x$ direction. The obtained average value of the Poisson's ratio is about -0.30. This result confirms the auxeticity of a finite sample of A$\gamma$G structure at room temperature. 
The results are numerically different from that of an infinite network because the boundary conditions on the structure (which opposes the lateral expansion) are different, and the interactions between the  A$\gamma$G structure and the substrate can influence its deformation mechanisms. Videos of these simulations (videos S3 and S4) are also shown in supplementary material.

%%%%%%%%%%%%%%%%%%%%%%%%%%%%%%%%%%%
%%% Conclusion
\section{Conclusions}

In summary, a new auxetic structure, based on the densest $\gamma$-graphyne structure, here named {\it A$\gamma$G structure}, is proposed and its mechanical and electronic properties analyzed using classical reactive and quantum  molecular dynamics simulations. We found that A$\gamma$G has a large bandgap of 2.48 eV, which will prevent electronic excitation coming from low-energy thermal or optical interactions. %%located at the M point. 
Its thermal stability was validated with the different adopted methods. From the mechanical point of view, it presents a low Young's modulus value when compared to that of graphene or even to that of $\gamma$-graphyne (about 300 and 163 GPa.nm~\cite{fonseca2017mestradosergio}, respectively). This result is expected since the A$\gamma$G structure consists of a $\gamma$-graphyne structure with several acetylenic chains/arms removed. Our results show that the A$\gamma$G is auxetic both when isolated (vacuum) and when deposited on a copper substrate. We believe that this  is the densest auxetic structure belonging to the graphyne-like families.  

\begin{section}{ACKNOWLEDGMENTS}
 Authors acknowledge support from the Brazilian agencies FAPESP (grants $\#$2013/08293-7, $\#$2018/03961-5 and $\#$2020/02044-9) and CNPq (grants $\#$437034/2018-6, $\#$310369/2017-7 and $\#$311587/2018-6). MJC, DSG, and TJS would like to thank the National Institute for Science and Technology on Organic Electronics (CNPq 573762/2008-2 and FAPESP 2008/57706-4). The authors acknowledge support from the John David Rogers Computing Center (CCJDR) at the Institute of Physics “Gleb Wataghin”, University of Campinas. Part of the computational resources was also provided by the National Laboratory for Scientific Computing, LNCC/MCTIC(LNCC/SantosDumont).

\end{section}

\footnotesize{
\bibliography{auxeticGraphyne} %your .bib file

\begin{thebibliography}{10}
\expandafter\ifx\csname url\endcsname\relax
  \def\url#1{\texttt{#1}}\fi
\expandafter\ifx\csname urlprefix\endcsname\relax\def\urlprefix{URL }\fi
\expandafter\ifx\csname href\endcsname\relax
  \def\href#1#2{#2} \def\path#1{#1}\fi

\bibitem{reviewEvans2000}
K.~E. Evans, K.~Alderson, Auxetic materials: the positive side of being
  negative, Engineering Science {\&} Education Journal 9 (2000) 148--154(6).
\newblock \href {http://dx.doi.org/10.1049/esej_20000402}
  {\path{doi:10.1049/esej_20000402}}.

\bibitem{review2Evans2000}
K.~E. Evans, A.~Alderson, Auxetic materials: Functional materials and
  structures from lateral thinking!, Advanced Materials 12~(9) (2000) 617--628.
\newblock \href
  {http://dx.doi.org/10.1002/(SICI)1521-4095(200005)12:9<617::AID-ADMA617>3.0.CO;2-3}
  {\path{doi:10.1002/(SICI)1521-4095(200005)12:9<617::AID-ADMA617>3.0.CO;2-3}}.

\bibitem{neville2013}
G.~N. Greaves, Poisson's ratio over two centuries: challenging hypotheses,
  Notes and Records: the Royal Society Journal of the History of Science 67~(1)
  (2013) 37--58.
\newblock \href {http://dx.doi.org/10.1098/rsnr.2012.0021}
  {\path{doi:10.1098/rsnr.2012.0021}}.

\bibitem{evans1991scicorrespondence}
K.~E. Evans, M.~A. Nkansah, I.~J. Hutchinson, S.~C. Rogers, Molecular network
  design, Nature 353~(6340) (1991) 124--124.
\newblock \href {http://dx.doi.org/10.1038/353124a0}
  {\path{doi:10.1038/353124a0}}.

\bibitem{lakes1987}
R.~Lakes, Foam structures with a negative {{P}}oisson{\textquoteright}s ratio,
  Science 235~(4792) (1987) 1038--1040.
\newblock \href {http://dx.doi.org/10.1126/science.235.4792.1038}
  {\path{doi:10.1126/science.235.4792.1038}}.

\bibitem{lakes1993}
R.~Lakes, Advances in negative {{P}}oisson's ratio materials, Advanced
  Materials 5~(4) (1993) 293--296.
\newblock \href {http://dx.doi.org//10.1002/adma.19930050416}
  {\path{doi:/10.1002/adma.19930050416}}.

\bibitem{raydouglas1993}
R.~H. Baughman, D.~S. Galv{\~{a}}o, Crystalline networks with unusual predicted
  mechanical and thermal properties, Nature 365~(6448) (1993) 735--737.
\newblock \href {http://dx.doi.org/10.1038/365735a0}
  {\path{doi:10.1038/365735a0}}.

\bibitem{QingAuxetichoneycomb2019}
W.~Wang, C.~He, L.~Xie, Q.~Peng, The temperature-sensitive anisotropic negative
  {{P}}oisson’s ratio of carbon honeycomb, Nanomaterials 9~(4) (2019) 487.
\newblock \href {http://dx.doi.org/10.3390/nano9040487}
  {\path{doi:10.3390/nano9040487}}.

\bibitem{silvaCORK2005}
S.~P. Silva, M.~A. Sabino, E.~M. Fernandes, V.~M. Correlo, L.~F. Boesel, R.~L.
  Reis, Cork: properties, capabilities and applications, International
  Materials Reviews 50~(6) (2005) 345--365.
\newblock \href {http://dx.doi.org/10.1179/174328005X41168}
  {\path{doi:10.1179/174328005X41168}}.

\bibitem{grima2011}
J.~N. Grima, D.~Attard, Molecular networks with a near zero
  {{P}}oisson{\textquoteright}s ratio, Physica Status Solidi (b) 248~(1) (2011)
  111--116.
\newblock \href {http://dx.doi.org/10.1002/pssb.201083979}
  {\path{doi:10.1002/pssb.201083979}}.

\bibitem{ray2015}
Y.~Wu, N.~Yi, L.~Huang, T.~Zhang, S.~Fang, H.~Chang, N.~Li, J.~Oh, J.~A. Lee,
  M.~Kozlov, A.~C. Chipara, H.~Terrones, P.~Xiao, G.~Long, Y.~Huang, F.~Zhang,
  L.~Zhang, X.~Lepr{\'{o}}, C.~Haines, M.~D. Lima, N.~P. Lopez, L.~P.
  Rajukumar, A.~L. Elias, S.~Feng, S.~J. Kim, N.~T. Narayanan, P.~M. Ajayan,
  M.~Terrones, A.~Aliev, P.~Chu, Z.~Zhang, R.~H. Baughman, Y.~Chen,
  Three-dimensionally bonded spongy graphene material with super compressive
  elasticity and near-zero {{P}}oisson{\textquoteright}s ratio, Nature
  Communications 6 (2015) 6141.
\newblock \href {http://dx.doi.org/10.1038/ncomms7141}
  {\path{doi:10.1038/ncomms7141}}.

\bibitem{fonseca2020}
V.~Gaal, V.~Rodrigues, S.~O. Dantas, D.~S. Galvão, A.~F. Fonseca, New zero
  poisson's ratio structures, Physica Status Solidi (RRL) – Rapid Research
  Letters 14~(3) (2020) 1900564.
\newblock \href {http://dx.doi.org/10.1002/pssr.201900564}
  {\path{doi:10.1002/pssr.201900564}}.

\bibitem{review2016}
J.-W. Jiang, S.~Y. Kim, H.~S. Park, Auxetic nanomaterials: Recent progress and
  future development, Applied Physics Reviews 3~(4) (2016) 041101.
\newblock \href {http://dx.doi.org/10.1063/1.4964479}
  {\path{doi:10.1063/1.4964479}}.

\bibitem{Mortazavi2017}
B.~Mortazavi, M.~Shahrokhi, M.~Makaremi, T.~Rabczuk, Anisotropic mechanical and
  optical response and negative poisson's ratio in {{M}}o2{{C}} nanomembranes
  revealed by first-principles simulations, Nanotechnology 28~(11) (2017)
  115705.
\newblock \href {http://dx.doi.org/10.1088/1361-6528/aa5c29}
  {\path{doi:10.1088/1361-6528/aa5c29}}.

\bibitem{gibs1982}
L.~J. Gibson, M.~F. Ashby, G.~S. Schajer, C.~I. Robertson, The mechanics of
  two-dimensional cellular materials, Proceedings of the Royal Society of
  London. A. Mathematical and Physical Sciences 382~(1782) (1982) 25--42.
\newblock \href {http://dx.doi.org/10.1098/rspa.1982.0087}
  {\path{doi:10.1098/rspa.1982.0087}}.

\bibitem{woj1987}
K.~Wojciechowski, Constant thermodynamic tension {{M}}onte {{C}}arlo studies of
  elastic properties of a two-dimensional system of hard cyclic hexamers,
  Molecular Physics 61~(5) (1987) 1247--1258.
\newblock \href {http://dx.doi.org/10.1080/00268978700101761}
  {\path{doi:10.1080/00268978700101761}}.

\bibitem{evans1994}
M.~A. Nkansah, K.~E. Evans, I.~J. Hutchinson, Modelling the mechanical
  properties of an auxetic molecular network, Modelling and Simulation in
  Materials Science and Engineering 2~(3) (1994) 337--352.
\newblock \href {http://dx.doi.org/10.1088/0965-0393/2/3/004}
  {\path{doi:10.1088/0965-0393/2/3/004}}.

\bibitem{evans1995}
K.~E. Evans, A.~Alderson, F.~R. Christian, Auxetic two-dimensional polymer
  networks. an example of tailoring geometry for specific mechanical
  properties, J. Chem. Soc.{,} Faraday Trans. 91 (1995) 2671--2680.
\newblock \href {http://dx.doi.org/10.1039/FT9959102671}
  {\path{doi:10.1039/FT9959102671}}.

\bibitem{griffin1998}
C.~He, P.~Liu, A.~C. Griffin, Toward negative {{P}}oisson ratio polymers
  through molecular design, Macromolecules 31~(9) (1998) 3145--3147.
\newblock \href {http://dx.doi.org/10.1021/ma970787m}
  {\path{doi:10.1021/ma970787m}}.

\bibitem{grimaevans2000}
J.~N. Grima, K.~E. Evans, Self expanding molecular networks, Chem. Commun.
  (2000) 1531--1532\href {http://dx.doi.org/10.1039/B004305M}
  {\path{doi:10.1039/B004305M}}.

\bibitem{grima2008}
J.~N. Grima, D.~Attard, R.~N. Cassar, L.~Farrugia, L.~Trapani, R.~Gatt, On the
  mechanical properties and auxetic potential of various organic networked
  polymers, Molecular Simulation 34~(10-15) (2008) 1149--1158.
\newblock \href {http://dx.doi.org/10.1080/08927020802512187}
  {\path{doi:10.1080/08927020802512187}}.

\bibitem{susuki2016}
Y.~Suzuki, G.~Cardone, D.~Restrepo, T.~S. Zavattieri, Pablo D.~Baker, F.~A.
  Tezcan, Self-assembly of coherently dynamic, auxetic, two-dimensional protein
  crystals, Nature 533~(7603) (2016) 369--373.
\newblock \href {http://dx.doi.org/10.1038/nature17633}
  {\path{doi:10.1038/nature17633}}.

\bibitem{LiACSOmega2020}
M.~Li, K.~Yuan, Y.~Zhao, Z.~Gao, X.~Zhao, A novel hyperbolic two-dimensional
  carbon material with an in-plane negative {{P}}oisson’s ratio behavior and
  low-gap semiconductor characteristics, ACS Omega 5~(26) (2020) 15783--15790.
\newblock \href {http://dx.doi.org/10.1021/acsomega.0c00182}
  {\path{doi:10.1021/acsomega.0c00182}}.

\bibitem{Baughman876687}
R.~H. Baughman, H.~Eckhardt, M.~Kertesz, Structure-property predictions for new
  planar forms of carbon: Layered phases containing sp2 and sp atoms, The
  Journal of Chemical Physics 87~(11) (1987) 6687--6699.
\newblock \href {http://dx.doi.org/10.1063/1.453405}
  {\path{doi:10.1063/1.453405}}.

\bibitem{ivanovsREVIEW2013}
A.~Ivanovskii, Graphynes and graphdyines, Progress in Solid State Chemistry
  41~(1) (2013) 1--19.
\newblock \href {http://dx.doi.org/10.1016/j.progsolidstchem.2012.12.001}
  {\path{doi:10.1016/j.progsolidstchem.2012.12.001}}.

\bibitem{liREVIEW2014}
Y.~Li, L.~Xu, H.~Liu, Y.~Li, Graphdiyne and graphyne: from theoretical
  predictions to practical construction, Chem. Soc. Rev. 43 (2014) 2572--2586.
\newblock \href {http://dx.doi.org/10.1039/C3CS60388A}
  {\path{doi:10.1039/C3CS60388A}}.

\bibitem{malko2008nonnullbandgapGYs}
D.~Malko, C.~Neiss, F.~Vi\~nes, A.~G\"orling, Competition for graphene:
  Graphynes with direction-dependent dirac cones, Phys. Rev. Lett. 108 (2012)
  086804.
\newblock \href {http://dx.doi.org/10.1103/PhysRevLett.108.086804}
  {\path{doi:10.1103/PhysRevLett.108.086804}}.

\bibitem{wangthermoeletric2013GYs}
X.-M. Wang, D.-C. Mo, S.-S. Lu, On the thermoelectric transport properties of
  graphyne by the first-principles method, The Journal of Chemical Physics
  138~(20) (2013) 204704.
\newblock \href {http://dx.doi.org/10.1063/1.4806069}
  {\path{doi:10.1063/1.4806069}}.

\bibitem{cevikthermoelectric2014GYs}
H.~Sevinçli, C.~Sevik, Electronic, phononic, and thermoelectric properties of
  graphyne sheets, Applied Physics Letters 105~(22) (2014) 223108.
\newblock \href {http://dx.doi.org/10.1063/1.4902920}
  {\path{doi:10.1063/1.4902920}}.

\bibitem{fonseca2017mestradosergio}
S.~A. Hernandez, A.~F. Fonseca, Anisotropic elastic modulus, high {{P}}oisson's
  ratio and negative thermal expansion of graphynes and graphdiynes, Diamond
  and Related Materials 77 (2017) 57--64.
\newblock \href {http://dx.doi.org/10.1016/j.diamond.2017.06.002}
  {\path{doi:10.1016/j.diamond.2017.06.002}}.

\bibitem{buheler2011GYsmech}
S.~W. Cranford, M.~J. Buehler, Mechanical properties of graphyne, Carbon
  49~(13) (2011) 4111--4121.
\newblock \href {http://dx.doi.org/10.1016/j.carbon.2011.05.024}
  {\path{doi:10.1016/j.carbon.2011.05.024}}.

\bibitem{nuno2021}
D.~Galhofo, N.~Silvestre, Computational simulation of $\gamma$-graphynes under
  monotonic and hysteretic loading, Mechanics of Advanced Materials and
  Structures 28~(5) (2021) 495--505.
\newblock \href {http://dx.doi.org/10.1080/15376494.2019.1578007}
  {\path{doi:10.1080/15376494.2019.1578007}}.

\bibitem{QPengGraphynes2020}
Y.~Yang, Q.~Cao, Y.~Gao, S.~Lei, S.~Liu, Q.~Peng, High impact resistance in
  graphyne, RSC Adv. 10 (2020) 1697--1703.
\newblock \href {http://dx.doi.org/10.1039/C9RA09685J}
  {\path{doi:10.1039/C9RA09685J}}.

\bibitem{plimpton1995}
S.~Plimpton, Fast parallel algorithms for short-range molecular dynamics,
  Journal of Computational Physics 117~(1) (1995) 1--19.
\newblock \href {http://dx.doi.org/10.1006/jcph.1995.1039}
  {\path{doi:10.1006/jcph.1995.1039}}.

\bibitem{andrewPRB2019}
R.~C. Andrew, R.~E. Mapasha, A.~M. Ukpong, N.~Chetty, Mechanical properties of
  graphene and boronitrene, Phys. Rev. B 85 (2012) 125428.
\newblock \href {http://dx.doi.org/10.1103/PhysRevB.85.125428}
  {\path{doi:10.1103/PhysRevB.85.125428}}.

\bibitem{brenner2002}
D.~W. Brenner, O.~A. Shenderova, J.~A. Harrison, S.~J. Stuart, B.~Ni, S.~B.
  Sinnott, A second-generation reactive empirical bond order ({REBO}) potential
  energy expression for hydrocarbons, Journal of Physics: Condensed Matter
  14~(4) (2002) 783--802.
\newblock \href {http://dx.doi.org/10.1088/0953-8984/14/4/312}
  {\path{doi:10.1088/0953-8984/14/4/312}}.

\bibitem{tao2013}
T.~Liang, T.-R. Shan, Y.-T. Cheng, B.~D. Devine, M.~Noordhoek, Y.~Li, Z.~Lu,
  S.~R. Phillpot, S.~B. Sinnott, Classical atomistic simulations of surfaces
  and heterogeneous interfaces with the charge-optimized many body {{(COMB)}}
  potentials, Materials Science and Engineering: R: Reports 74~(9) (2013)
  255--279.
\newblock \href {http://dx.doi.org/10.1016/j.mser.2013.07.001}
  {\path{doi:10.1016/j.mser.2013.07.001}}.

\bibitem{vanduin2001}
A.~C.~T. van Duin, S.~Dasgupta, F.~Lorant, W.~A. Goddard, Reaxff: A reactive
  force field for hydrocarbons, The Journal of Physical Chemistry A 105~(41)
  (2001) 9396--9409.
\newblock \href {http://dx.doi.org/10.1021/jp004368u}
  {\path{doi:10.1021/jp004368u}}.

\bibitem{mueller2010}
J.~E. Mueller, A.~C.~T. van Duin, W.~A. Goddard, Development and validation of
  {{R}}eaxff reactive force field for hydrocarbon chemistry catalyzed by
  nickel, The Journal of Physical Chemistry C 114~(11) (2010) 4939--4949.
\newblock \href {http://dx.doi.org/10.1021/jp9035056}
  {\path{doi:10.1021/jp9035056}}.

\bibitem{dftb1}
D.~Porezag, T.~Frauenheim, T.~Kohler, G.~Seifert, R.~Kaschner, Construction of
  tight-binding-like potentials on the basis of density-functional theory:
  Application to carbon, Physical Review B 51 (1995) 12947.

\bibitem{dftb2}
G.~Seifert, D.~Porezag, T.~Frauenheim, Calculations of molecules, clusters, and
  solids with a simplified {{LCAO-DFT-LDA}} scheme, Int. J. Quantum Chemistry
  58 (1996) 185.

\bibitem{Manzano123239}
H.~Manzano, A.~N. Enyashin, J.~S. Dolado, A.~Ayuela, J.~Frenzel, G.~Seifert, Do
  cement nanotubes exist?, Advanced Materials 24~(24) (2012) 3239--3245.
\newblock \href {http://dx.doi.org/10.1002/adma.201103704}
  {\path{doi:10.1002/adma.201103704}}.

\bibitem{koskinen2009density}
P.~Koskinen, V.~M{\"a}kinen, Density-functional tight-binding for beginners,
  Computational Materials Science 47~(1) (2009) 237--253.

\bibitem{Elstner987260}
M.~Elstner, D.~Porezag, G.~Jungnickel, J.~Elsner, M.~Haugk, T.~Frauenheim,
  S.~Suhai, G.~Seifert, Self-consistent-charge density-functional tight-binding
  method for simulations of complex materials properties, Physical Review B
  58~(11) (1998) 7260--7268.
\newblock \href {http://dx.doi.org/10.1103/PhysRevB.58.7260}
  {\path{doi:10.1103/PhysRevB.58.7260}}.

\bibitem{dftb+}
B.~Aradi, B.~Hourahine, T.~Frauenheim, {{DFTB+}}, a sparse matrix-based
  implementation of the {{DFTB}} method, The Journal of Physical Chemistry A
  111~(26) (2007) 5678--5684.
\newblock \href {http://dx.doi.org/10.1021/jp070186p}
  {\path{doi:10.1021/jp070186p}}.

\bibitem{Gemming2010}
S.~Gemming, A.~N. Enyashin, J.~Frenzel, G.~Seifert, Adsorption of nucleotides
  on the rutile (110) surface, International Journal of Materials Research
  101~(6) (2010) 758--764.
\newblock \href {http://dx.doi.org/10.3139/146.110337}
  {\path{doi:10.3139/146.110337}}.

\bibitem{Lukose2010}
B.~Lukose, A.~Kuc, J.~Frenzel, T.~Heine, On the reticular construction concept
  of covalent organic frameworks, Beilstein Journal of Nanotechnology 1 (2010)
  60--70.
\newblock \href {http://dx.doi.org/10.3762/bjnano.1.8}
  {\path{doi:10.3762/bjnano.1.8}}.

\bibitem{Kubar2013}
T.~Kubar, Z.~Bodrog, M.~Gaus, C.~Koehler, B.~Aradi, T.~Frauenheim, M.~Elstner,
  Parametrization of the {{SCC-DFTB}} method for halogens, Journal of Chemical
  Theory and Computation 9~(7) (2013) 2939--2949.
\newblock \href {http://dx.doi.org/10.1021/ct4001922}
  {\path{doi:10.1021/ct4001922}}.

\bibitem{evans1985nose}
D.~J. Evans, B.~L. Holian, The nose--hoover thermostat, The Journal of chemical
  physics 83~(8) (1985) 4069--4074.

\bibitem{vmd}
W.~Humphrey, A.~Dalke, K.~Schulten, {VMD}: Visual molecular dynamics, Journal
  of Molecular Graphics 14~(1) (1996) 33--38.
\newblock \href {http://dx.doi.org/10.1016/0263-7855(96)00018-5}
  {\path{doi:10.1016/0263-7855(96)00018-5}}.

\end{thebibliography}
}

\end{document}